\documentclass[aps,twocolumn,superscriptaddress, longbibliography,bibnotes]{revtex4-1}
\usepackage{amsmath}
\usepackage{float}
\usepackage{comment}
\usepackage{color}
\usepackage{physics}
\usepackage{graphicx}% Include figure files
\usepackage{dcolumn}% Align table columns on decimal point
\usepackage{bm}% bold math
\usepackage{todonotes}
\usepackage[colorlinks=true,bookmarks=false,citecolor=blue,linkcolor=blue,hyperfootnotes=true,urlcolor=blue]{hyperref}

\def\pipi{\ensuremath{(0.5, 0.5)}}
\def\hpihpi{\ensuremath{(0.25, 0.25)}}
\def\piz{\ensuremath{(0.5, 0)}}
\def\seven{Sr$_3$Ir$_2$O$_7$}
\def\four{Sr$_2$IrO$_4$}

\begin{document}

\title{Magnetism in artificial Ruddlesden-Popper iridates leveraged by structural distortions}

\author{D. Meyers}
\email{dmeyers@bnl.gov}
\author{Yue Cao}
\author{G. Fabbris}
\author{Neil~J.~Robinson} %SWT
\affiliation{Condensed Matter Physics and Materials Science Department, Brookhaven National Laboratory, Upton, New York 11973, USA}

\author{Lin Hao} 
\author{C.~Frederick}
\author{N.~Traynor}
\author{J.~Yang}
\affiliation{Department of Physics and Astronomy, University of Tennessee, Knoxville, Tennessee 37996, USA}

\author{Jiaqi Lin}
\affiliation{Beijing National Laboratory for Condensed Matter Physics and Institute of Physics, Chinese Academy of Sciences, Beijing 100190, China}
\affiliation{School of Physical Sciences, University of Chinese Academy of Sciences, Beijing 100049, China}

\author{M. H. Upton}
\author{D. Casa}
\author{Jong-Woo Kim}
\author{T. Gog}
\author{E. Karapetrova}
\author{Yongseong~Choi}
\author{D. Haskel}
\affiliation{Advanced Photon Source, Argonne National Laboratory, Argonne, Illinois 60439, USA}
\author{P. J. Ryan}
\affiliation{Advanced Photon Source, Argonne National Laboratory, Argonne, Illinois 60439, USA}
\affiliation{School of Physical Sciences, Dublin City University, Dublin 9, Ireland}

\author{Lukas Horak}
\affiliation{Department of Condensed Matter Physics, Charles University, Ke Karlovu 3, Prague 12116, Czech Republic}

\author{X. Liu}
\affiliation{Beijing National Laboratory for Condensed Matter Physics and Institute of Physics, Chinese Academy of Sciences, Beijing 100190, China}
\affiliation{Collaborative Innovation Center of Quantum Matter, Beijing, China}

\author{Jian Liu}
\email{jianliu@utk.edu}
\affiliation{Department of Physics and Astronomy, University of Tennessee, Knoxville, Tennessee 37996, USA}
\author{M. P. M. Dean}
\email{mdean@bnl.gov}
\affiliation{Condensed Matter Physics and Materials Science Department, Brookhaven National Laboratory, Upton, New York 11973, USA}

\date{\today}

\begin{abstract}
We report on the tuning of magnetic interactions in superlattices composed of single and bilayer SrIrO$_3$ inter-spaced with SrTiO$_3$. Magnetic scattering shows predominately $c$-axis antiferromagnetic orientation of the magnetic moments for the bilayer justifying these systems as viable artificial analogues of the bulk Ruddlesden-Popper series iridates.  Magnon gaps are observed in both superlattices, with the magnitude of the gap in the bilayer being reduced to nearly half that in its bulk structural analogue, Sr$_3$Ir$_2$O$_7$. We assign this to modifications in the anisotropic exchange driven by bending of the $c$-axis Ir-O-Ir bond and subsequent local environment changes, as detected by x-ray diffraction and modeled using spin wave theory. These findings explain how even subtle structural modulations driven by heterostructuring in iridates are leveraged by spin orbit coupling to drive large changes in the magnetic interactions.
\end{abstract}

\pacs{}% PACS, the Physics and Astronomy
\maketitle

\begin{figure}
\begin{centering}
\includegraphics[width=0.5\textwidth]{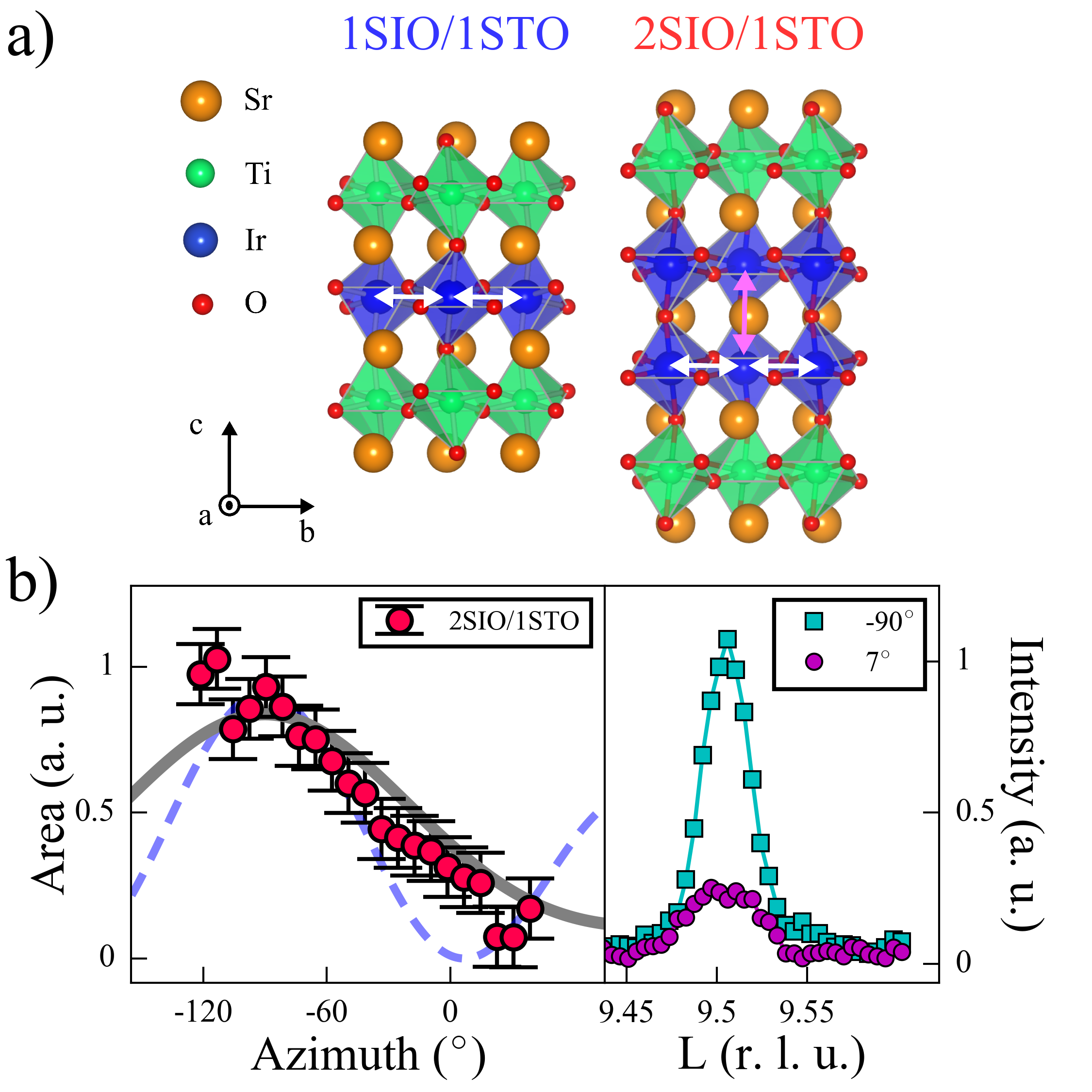} %
\caption{a) Structure of the two SLs. White arrows show in-plane magnetic exchange pathways and the pink arrow shows the out-of-plane exchange in the bilayer. b) Left panel: Integrated intensities as a function of the azimuth angle for the $(0.5, 0.5, 9.5)$ magnetic reflection of the 2SIO/1STO sample are shown as red circles, along with the expected response for $c$-axis (solid grey line) and in-plane moment orientations (blue dashed lines). Right panel: Normalized magnetic Bragg peaks at maximum and where no intensity is expected for in-plane moments.
\label{REXS}
}
\end{centering}
\end{figure}

%%%%%%%%%%%%%%%%%%%%%%%%%%%%%%%%
\section{Introduction}
%%%%%%%%%%%%%%%%%%%%%%%%%%%%%%%%
Recent years have seen iridates, compounds composed of active Ir $5d$ orbitals in oxygen octahedra, emerge as an important new class of strongly correlated materials \cite{Crawford1994_214mag,kim2008novel,kim2009phase,Mitchell2015_214, Witczak_correlatedSOC2014}. The combination of crystal field interactions and strong spin-orbit coupling generates  narrow bands leading to insulating antiferromagnetic ground states that arise from modest values of the Coulomb repulsion $U$ \cite{kim2008novel}. Many appealing structural and electronic analogies between iridates and lighter $3d$-electron based cuprates have been identified \cite{Crawford1994_214mag, wang2011twisted, kim2014fermi, cao2014hallmarks,Kim2016_dgapIr, liu2016anisotropic}. One crucial difference, however, is that iridates host spin-orbit coupled $J_\text{eff}=\frac{1}{2}$ magnetic moments, which have a more intricate coupling to orbital distortions than pure spin $S=\frac{1}{2}$ moments \cite{jackeli2009mott}. This is borne out in observations: different magnetic ground states appear in iridates composed of similar Ir-O octahedra when they are subtly distorted or interspaced with different atoms. Sr$_2$IrO$_4$, hosting isolated IrO$_2$ layers, forms an $ab$-plane canted antiferromagnetic state \cite{kim2009phase, cao1998_214}, while Sr$_3$Ir$_2$O$_7$, hosting isolated IrO$_2$ \textit{bi}-layers, has $c$-axis collinear antiferromagnetic ordering \cite{Cao2002_327structure, Kim2012_327REXS}.  Towards understanding how the structural modulations tailor magnetic ground states, resonant inelastic x-ray scattering (RIXS) has been extremely successful in quantifying the magnetic interactions present in iridate crystals \cite{kim2012magnetic, Kim2012_327RIXS, liu2012testing, Yin2013Ferromagnetic, kim2014excitonic, liu2016anisotropic}. For example, a large (92~meV) spin gap was measured in \seven{}, reflecting the substantial interlayer anisotropic coupling that causes a ``dimensionality driven spin flop transition" with respect to \four{} \cite{Kim2012_327RIXS, Sala2015_327DimerRIXS}. 

Recently, artificial layered iridates in analogy to Ruddlesden-Popper iridates were realized via alternating layers of $n$SrIrO$_3$ and SrTiO$_3$ ($n$SIO/1STO), which showed a metal-insulator transition as a function of $n$, closely mirroring their bulk analogues  \cite{Matsuno2015_SIOSTO,Kim2016_OptDFT}. However, an $ab$-plane canted antiferromagnetic state was argued to be maintained for $n\leq3$, suggesting that the spin flop transition is suppressed and breaking the analogy to bulk crystals  \cite{Matsuno2015_SIOSTO}. To achieve this change in the ground state, anisotropic coupling between the two Ir layers in 2SIO/1STO, which favors the $c$-axis antiferromagnetic state, would need to be substantially modified compared to \seven{} \cite{Kim2012_327REXS}. Further, this behavior is disputed by density functional theory (DFT) predictions that find $c$-axis collinear magnetism in 2SIO/1STO \cite{Kim2017_STSIOtheory,Kim2017_327theory}. Taken together, these conflicting results point to the need for direct observation of the exchange coupling and the resulting magnetic structure to unravel the impact of heterostructuring on the behavior of these proposed artificial analogues to the Ruddlesden-Popper iridates. 

In this work, we directly probe the magnetic behavior of $n$SIO/1STO and extend the sensitivity of Ir $L_3$ RIXS to quantify the interactions that stabilize this state. We find a $c$-axis antiferromagnetic ground state in 2SIO/1STO, in contrast with an earlier report \cite{Matsuno2015_SIOSTO}, demonstrating that the magnetic ground state mimics bulk \seven{}. In both 1SIO/1STO and 2SIO/1STO, the magnetic excitation spectrum shows a clear dispersion with a magnon gap of 55~meV in $n=2$, substantially reduced to about half that in bulk \seven{} \cite{Kim2012_327RIXS,Sala2015_327DimerRIXS}. Based on modeling the magnetic dispersion, the predominately $c$-axis moments in 2SIO/1STO are stabilized by the anisotropic coupling between Ir-O planes as seen in \seven{}. However, the lowering of the magnon gap evidences a significant reduction in the tetragonal distortion of the octahedra, while for 1SIO/1STO the gap size is similar to bulk \four{} \cite{pincini2017_214gap}. The source of the modulation of the tetragonal distortion was determined to be substantial bending of the $c$-axis Ir-O-Ir bonds alongside changes in the local environment beyond the octahedra. This work establishes these artificial structures as true analogues to the Ruddlesden-Popper iridates, with the caveat that changes in the magnetic ground states are highly susceptible to subtle structural distortions \cite{Bogdanov_214orbital2015}.  These distortions push the heterostructure towards a quantum critical point between the $ab$-plane and $c$-axis antiferromagnets, exemplifying how magnetic states in iridates can be transformed in a tractable manner owing to their strong spin-orbit coupling.

%%%%%%%%%%%%%%%%%%%%%%%%%%%%%%%%
\section{Experimental details}
%%%%%%%%%%%%%%%%%%%%%%%%%%%%%%%%

Superlattices (SLs) of form [$n$SIO/1STO]$\times m$ with $n = 1$, $2$ and $m = 60$, $30$, respectively, were grown with pulsed laser deposition using methods described in Ref.~\cite{Hao_arxiv}, as depicted in Fig.~\ref{REXS}(a). High sample quality was verified by x-ray diffraction, x-ray magnetic circular dichroism, transport and magnetometry measurements (Fig.~S1-S3 of~\cite{supplemental}), consistent with previous studies \cite{Hao_arxiv, Matsuno2015_SIOSTO}. Resonant elastic x-ray scattering (REXS), RIXS, and non-resonant diffraction data were taken at the 6-ID-B, 27-ID-B, and 33-BM-C beamlines of the Advanced Photon Source at Argonne National Laboratory. The RIXS energy resolution was 35~meV, full width half maximum. Further details are available in the Supplemental Material \cite{supplemental}.

%%%%%%%%%%%%%%%%%%%%%%%%%%%%%%%%
\section{Crystal and magnetic structure}
%%%%%%%%%%%%%%%%%%%%%%%%%%%%%%%%

\begin{figure}
\includegraphics[width=0.48\textwidth]{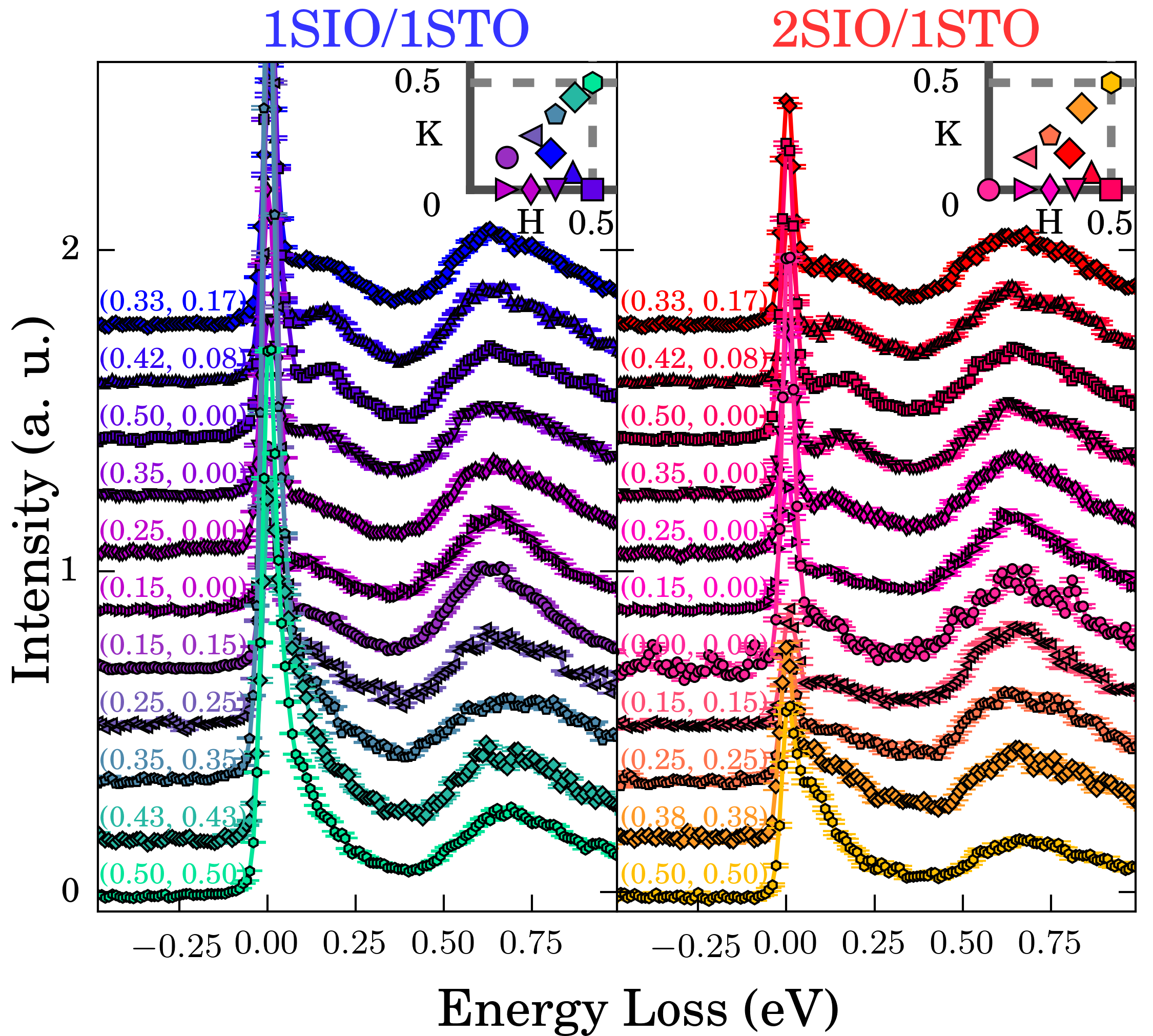} %
\caption{Raw RIXS spectra for 1SIO/1STO (left panel) and 2SIO/1STO (right panel), with measured $Q$-points shown in the inset, where the grey line denotes the Brillouin zone boundary.
\label{Spectra}
}
\end{figure}

Previous investigations of $n$SIO/1STO with $n = 1$, $2$, $3$ displayed a net ferromagnetic moment for all samples, which was taken as evidence for the stabilization of canted $ab$-plane magnetic moments as in \four{} \cite{Matsuno2015_SIOSTO}. Although there is a strong consensus that this is valid for 1SIO/1STO, the result for 2SIO/1STO is more controversial as it  breaks the analogy between $n=2$ and \seven{}, which has purely $c$-axis collinear antiferromagnetism implying no spontaneous net moment \cite{Kim2012_327REXS, Matsuno2015_SIOSTO}. Theory also predicted $c$-axis moments and posited that the observed net ferromagnetic moment comes from oxygen vacancies \cite{Kim2017_STSIOtheory, Kim2017_327theory,Nagai_327canted}. Establishing the true magnetic ground state is of high importance towards extracting the magnetic exchange parameters that ultimately dictate the overall magnetic behavior of these heterostructures.   

\begin{figure*}
\begin{centering}
\includegraphics[width=0.9\textwidth]{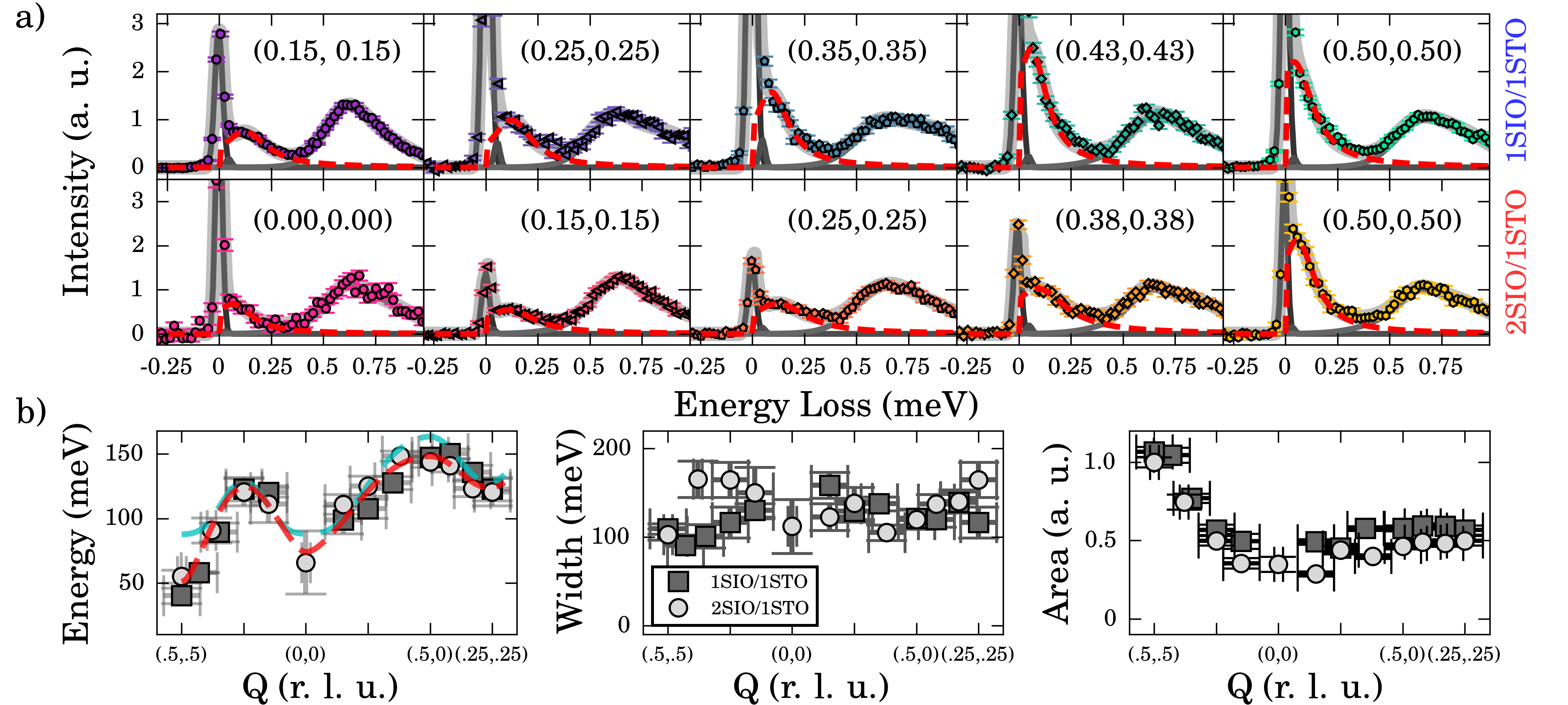} %
\caption{a) RIXS spectra for 1SIO/1STO (top) and 2SIO/1STO (bottom). Spectra colors follow from Fig.~\ref{Spectra}. The magnon feature is displayed as a dashed red line, while other contributions are in various greys, described in text. b) Extracted energy loss, width, and integrated intensity, respectively, of the magnon feature across reciprocal space. The dashed red (cyan) line is the magnon dispersion fit for 2SIO/1STO (\seven{}). The errors are statistical from the fitting, with the errors carried through the area integration calculation. 
\label{Fit} }
\end{centering}
\end{figure*}

In view of this controversy, we directly measured the spin ordering direction using azimuthal REXS scans, as was done in \seven{} \cite{Kim2012_327RIXS,Boseggia2012_327REXS}. This dependence is shown for 2SIO/1STO in Fig.~\ref{REXS}(b), left panel  \footnote{For the magnetic reflection the SL structure is used for r. l. u, with $a\approx b \approx c/3$.}.  The calculated azimuthal dependence for $c$-axis oriented antiferromagnetic moments, shown as the grey line, matches the data well and establishes predominately $c$-axis moments \cite{supplemental}. To further emphasize this distinction, we also show the magnetic Bragg peaks where the maximum intensity for both cases is expected (-90$^{\circ}$), and also where no intensity for in-plane moments is expected (7$^{\circ}$), Fig. \ref{REXS} (b), right panel. Clearly, a magnetic peak persists with integrated intensity that matches that expected for $c$-axis orient moments ($\sim~30$\%). These results then agree with theoretical predictions, showing the 2SIO/1STO SL maintains the same magnetic ground state as \seven{}, strengthening the analogy to Ruddlesden-Popper series iridates \cite{Kim2017_STSIOtheory}. Armed with the correct ground state, the magnetic excitation spectrum can be correctly interpreted, allowing the probing of the magnetic exchange couplings to unravel the true impact of artificial heterostructuring.

To directly probe the magnetic interactions, we utilize RIXS to map the magnetic dispersion of the established magnetic ground state. Although Ir $L_3$-edge RIXS has been applied extensively to iridate crystals and thin films, a full 2D magnetic dispersion curve has never been characterized on SL heterostructures due to the relatively large (5~$\mathrm{\mu}$m) x-ray penetration depth at the Ir L$_3$ edge
\cite{kim2012magnetic,Kim2012_327RIXS,Yue2017_Dop214RIXS,Sala2015_327DimerRIXS,pincini2017_214gap,Lupascu2014strain214,liu2016anisotropic,Gretarsson2016_dop214RIXS,gruenewald2017engineering}. This challenge was overcome by growing  relatively thick SLs (60 IrO$_2$ planes) and working near grazing incidence (1$^{\circ}$) \cite{supplemental}. Raw RIXS spectra for the SLs are displayed in Fig.~\ref{Spectra}. Each spectra displays a high energy feature around 0.75 eV energy loss, corresponding to both an intra-$t_{2g}$ orbital excitation and the $e-h$ continuum \cite{kim2014excitonic}. A sharp peak arises at zero energy due to elastic scattering, along with a small phonon feature at around 40 meV. Finally, a dispersive feature from 50 to 140 meV is seen in all spectra, and is identified as the magnon excitation \footnote{Only one magnetic excitation was observed for both samples, despite the presence of both optical and acoustic modes for the bilayer. This is due to the intensity dependence of each mode as discussed in section IV. }, with the higher energy tail including multimagnon excitations \cite{kim2012magnetic,Kim2012_327RIXS,Yue2017_Dop214RIXS,liu2016anisotropic,Gretarsson2016_dop214RIXS}. The spectra were fit using a combination of peaks in a similar approach to that used previously (see supplemental materials)  \cite{Kim2012_327RIXS,kim2014excitonic,Yue2017_Dop214RIXS,pincini2017_214gap,supplemental}. Examples of fits along the nodal direction for each sample are displayed in Fig.~\ref{Fit}(a). From these one can extract the energy, width, and integrated intensity of the magnetic excitation, Figs.~\ref{Fit}(b). The intensity peaks at the magnetic ordering wave vector \pipi{} and the energy loss is within the bandwidth seen for \four{} and \seven{}, corroborating our assignment of the feature as a magnetic type excitation \cite{kim2012magnetic,Kim2012_327RIXS}. 

From the extracted magnon dispersion, some important observations are immediately clear: (i) both SLs have nearly identical dispersion around the \hpihpi{} and \piz{} points with maxima of $\sim$ 120 and 150 meV, respectively, (ii) both samples show magnon gaps. In the case of the 1SIO/1STO, the size of the gap is not well defined, being $\sim11-36$ meV, due to the worse reciprocal space resolution, 0.46~\AA{}$^{-1}$ (0.073~r.l.u.) \cite{supplemental}. For 2SIO/1STO, a mask was used to improve the resolution to 0.12 \AA{}$^{-1}$ (0.018~r.l.u.) for the \pipi{} and (0, 0) Q-points. Here, a larger gap is much better defined as between 50 and 60~meV at \pipi{} \cite{supplemental}. Compared with the dispersion for \seven{}, Fig.~\ref{Fit}(b), the overlap is very robust everywhere \textit{except} at these minima \cite{Kim2012_327RIXS}.

\begin{figure}
\begin{centering}
\includegraphics[width=0.425\textwidth]{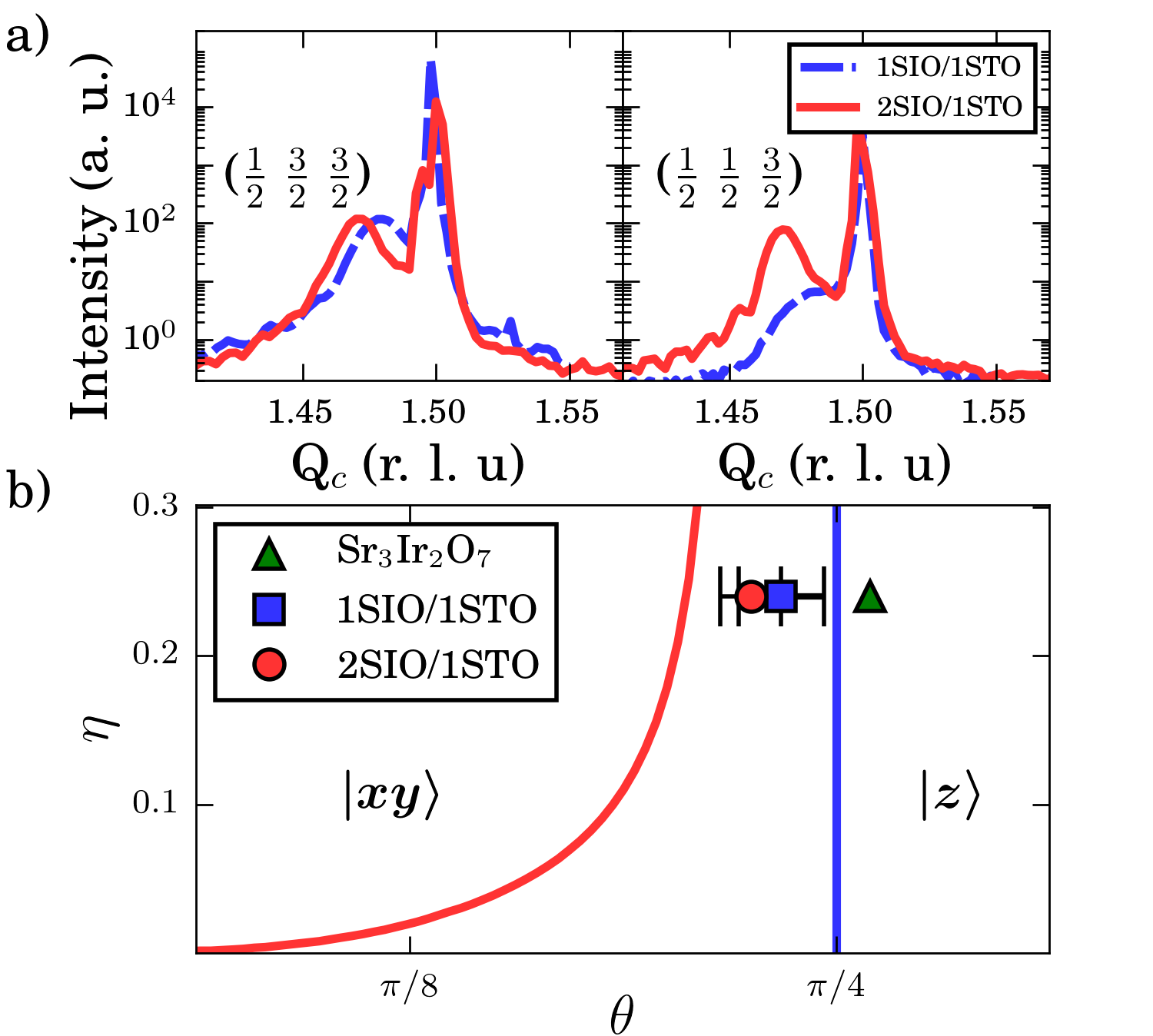}
\caption{a) Rotation and tilt peaks, left and right panels, for $n=1,~2$ SLs. Sharp features at Q$_c$ = 1.5 are substrate reflections, while broad features are sample reflections. b) Classical phase diagram as a function of $\eta$ and $\theta$. Phase boundaries between the canted in-plane $|xy\rangle$ and collinear out-of-plane $|z\rangle$ orderings are displayed for single layer (blue) and bilayer iridates (red). The position for \seven{} (green triangle) was taken from~\cite{Kim2012_327RIXS}. Error bars are the statistical error (magnon gap range) of $\theta$ for 2SIO/1STO (1SIO/1STO).}
\label{RotateModel}
\end{centering}
\end{figure}

We analyzed the origin of this anomalous behavior using linear spin wave theory \footnote{It should be noted, for \seven{} the dispersion and magnon gap can also be described in terms of a bond-operator approach utilizing spin dimers along the $c$-axis \cite{Sala2015_327DimerRIXS}, not considered here for several reasons. In this case, the SLs lies firmly within the antiferromagnetic order regime as clearly evidenced by the smaller magnon gaps, whereas the dimer model requires gaps $\geq$ 90~meV. Finally, the sizable total moments observed, on the order of \four{}, also conflict with a dimer picture in this case \cite{supplemental}.} applied to the Hamiltonian described in Ref.~\cite{Kim2012_327RIXS}, with $ab$-plane canted and $c$-axis collinear ground states for the single and bilayer SLs, respectively \cite{Kim2012_327RIXS,supplemental}. Importantly, the canted nature of the moments in 1SIO/1STO gives a dispersion relation that fundamentally differs from the out-of-plane N\'eel state \cite{Kim2012_327RIXS,supplemental}. As was the case for \seven{}, the nine magnetic couplings can be parameterized in terms of: (i) the tetragonal distortion $\theta$, defined by $\tan{2\theta} = \frac{2\sqrt{2}\lambda}{\lambda - 2\Delta}$, with spin-orbit coupling $\lambda$ and tetragonal splitting $\Delta$, (ii) $\eta=\frac{J_H}{U}$, with Hund's coupling $J_H$ and Coulomb repulsion $U$, (iii) the octahedral rotation angle $\alpha$. The form of the exchange couplings in terms of these parameters is described in the supplemental ($J^{\prime}_{ab}$, $J^{\prime\prime}_{ab}$, and  $J^{\prime}_c$ are treated as free fit parameters) \cite{Kim2012_327REXS, Kim2012_327RIXS, supplemental}. Concerning (ii), $\eta = 0.24$ was established for \seven{} and is unlikely to change significantly, leaving the tetragonal distortion and rotation angles to explain the observed dispersion and magnon gaps \footnote{As shown in Fig. \ref{RotateModel}(b), the phase transition is not sensitive to $\eta$ in this regime, nor is the dispersion fit.}.

Regarding $\alpha$, \four{} and \seven{} both feature large in-plane rotations ($\alpha\,=\,12^{\circ}$ and $11^{\circ}$ respectively), but no tilts (rotations about $a$/$b$ axes that bend the $c$-axis bond) \cite{Crawford1994_214mag,Cao2002_327structure,Hogan_327refinement2016}. Bulk-like SIO films, on the other hand, show substantial tilts and rotations implying that similar effects may be present in SLs \cite{Liu2016_SIO}. We consequently tested for the presence of octehedral tilts and rotations by scanning the half order Bragg peaks locations. While an exact structural solution of the SLs is unfeasible due to the complex orthorhombic structure of SIO \cite{Liu2016_SIO}, published methods allow us to associate different half order reflections with different antiphase distortions \cite{May2010, Brahlek2017,glazer1975}. We measured several reflections for the $n=1$ and $2$ samples and illustrate the important behavior in Fig.~\ref{RotateModel}(a) \cite{supplemental}. The $(\frac{1}{2}, \frac{3}{2}, \frac{3}{2})$ reflection (left panel) arises from a combination of rotations and tilts; whereas the $(\frac{1}{2}, \frac{1}{2}, \frac{3}{2})$ reflection (right panel) comes from \emph{only} tilts. Both peaks are of similar magnitude for $n=2$, but the tilt-peak is suppressed by an order of magnitude in $n=1$. This data suggests that both SLs have similar rotations of $\sim$ 8$^{\circ}$. In contrast with the nearly straight $c$-axis bonds seen in \seven{} \cite{Cao2002_327structure}, $n=2$ likely hosts tilting of a similar size (of order 8$^{\circ}$), while $n=1$ has small, but finite tilts  \footnote{In fitting the magnetic dispersion, values of $\alpha$ in the range $5-15^{\circ}$ are consistent with the data and thus we fix $\alpha~=~8^{\circ}$}. Most importantly, the presence of tilting generates the $ab$-plane ferromagnetic moment in 2SIO/1STO, observed experimentally \cite{Matsuno2015_SIOSTO,Hao_arxiv}, through canting the $c$-axis antiferromagnetic moments, resolving the conflict between previous experimental interpretations and theory \cite{Kim2017_STSIOtheory,Kim2017_327theory, vacancies}.

%%%%%%%%%%%%%%%%%%%%%%%%%%%%%%%%%%%%
\section{Tuning magnetic ground states}
%%%%%%%%%%%%%%%%%%%%%%%%%%%%%%%%%%%%

\begin{table}[t]
\begin{centering}
\begin{tabular}{| m{0.6cm}|| m{0.7cm} | m{0.7cm} | m{0.7cm} | m{0.7cm} | m{0.7cm} | m{0.7cm} | m{0.7cm}| m{0.7cm}| m{0.7cm}|}
\hline
  $n$ &  $J_{ab}$  &  $J_c$ &   $\Gamma_{ab}$  &   $\Gamma_c$   &   $D_{ab}$   &   $D_c$ & $J^{\prime}_{ab}$ & $J^{\prime\prime}_{ab}$ & $J^{\prime}_c$   \\ \hline
1 & 57.2   & ~ - ~ &   -0.2  &  ~ - ~  &  11.9   &  ~ - ~ &   -14.8 &   1.5  &  ~ - ~   \\ \hline
2 & 73.0   &  41.1  &   -1.5  &   25.2    &   16.1   &   25.7 &   4.1  &  8.3 &   3.4    \\ \hline
\end{tabular}
\caption{Exchange parameters (in meV) for $n$SIO/1STO from the best fit of tetragonal distortion $\theta$. $J$, $\Gamma$ and $D$ are the Heisenberg, anisotropic and Dzyaloshinskii-Moriya interactions respectively. Subscripts denote coupling directions and $^{\prime}$ indicates next neighbor coupling.}
\label{couplings}
\end{centering}
\end{table}

Having established the approximate rotation angles of both samples as $\alpha=8^{\circ}$, we can fit the dispersion of the SLs, as discussed above, displayed for 2SIO/1STO in Fig.~\ref{Fit}(b) left panel \cite{supplemental}. For 1SIO/1STO with gap $\sim11-36$ meV (compared to $\sim$25~meV for \four{}), we find $\theta=0.221$-$0.247\pi$. Within this range, spin wave theory can model well the dispersion throughout reciprocal space \cite{pincini2017_214gap}. For 2SIO/1STO, fitting with $\theta=0.225\pm0.009 \pi$  adequately reproduces the gaps and dispersion. For the bilayer, optical and acoustic modes are present, but only one mode is observed due to the $Q$-dependence of their intensities, discussed in the supplemental \cite{supplemental}. The similar theta values, which do not reflect the differences seen between bulk \four{} and \seven{}, are initially surprising due to the less distorted octahedra observed for both SIO and \seven{} ($\leq$ 2\%, $\sim$ 8\% for \four{}) \cite{Liu2016_SIO,Cao2002_327structure,Hogan_327refinement2016,Bogdanov_214orbital2015,Kim2012_327RIXS}. However, the Ir $5d$ orbitals are rather extended spatially and couple strongly to next nearest neighbors, breaking the local symmetry \cite{Bogdanov_214orbital2015}. The similar $\theta$ of the SLs further corroborates this, pointing to the SL structure as the dominant determinant of $\theta$.

Extracted values of the exchange couplings for each of the SLs are shown in Table ~\ref{couplings} \cite{supplemental}. For 1SIO/1STO, the values are similar to those found in both doped and undoped \four{}, owing to the relatively small change in $\theta$ between the SL and \four{} ($\sim$ 0.01$\pi$) \cite{pincini2017_214gap}. This indicates the 1SIO/1STO provides another magnetic analogue to cuprates, similar to that found in \four{}, but with the higher tunability afforded by heterostructuring \cite{gao_214SC_2015,yan2015electron,kim2016observation,Mitchell2015_214}. Comparing 2SIO/1STO with \seven{}, on the other hand, the changes are quite substantial, owing to the much more significant shift of $\theta$ (0.035$\pi$) \cite{Kim2012_327RIXS}. Here, the ratio $J_{ab}/J_{c}$ is half that found in \seven{}, a reasonable change in light of the more uniform octahedra expected in the SL (due to the ability of tilting to circumvent distortion) \footnote{Within the dimer model, this ratio is strongly inverted with much larger $J_c$ and smaller $J_{ab}$ \cite{Sala2015_327DimerRIXS}.}. Finally, from fitting the smaller magnon gap,  a 28\% decrease in the pseudodipolar anisotropic coupling $\Gamma_c$ is observed, which is chiefly responsible for stabilizing the $c$-axis magnetic ground state.

To investigate the stability of the observed magnetic phases, we map the SLs on the classical ($\theta,\eta$) magnetic phase diagram in Fig.~\ref{RotateModel}(b) alongside \four{} and \seven{} \cite{Kim2012_327REXS}. Intriguingly, both SLs lie close to their respective phase transitions. This is especially significant for 2SIO/1STO, where the material lies closer to the phase transition than its bulk analogue, within $0.02\pi$ or $\sim$ 10 meV tetragonal splitting change. The fact that such a shift happens despite the similar structures of 2SIO/1STO and \seven{} shows how relatively small distortions can strongly modify the interactions of these SLs. 
Further bending of the $c$-axis bond can then be expected to drive the system closer to, and eventually through, a quantum critical point between the $ab$-plane canted and $c$-axis collinear antiferromagnets. This could be accomplished through applying epitaxial tensile strain, by changing the substrate, or by substituting Sr with Ca. Based on the calculated change in the crystal field of strained films of \four{}, applied strain of only a few percent could be enough to drive 2SIO/1SIO to the quantum critical point \cite{Lupascu2014strain214,Kim2017_STSIOtheory}. In this way, strong spin-orbit coupling provides a means to exploit small structural distortions to stabilize large changes in the magnetic ground state.

%%%%%%%%%%%%%%%%%%%%%%%%%%%%%%%%%%%%
\section{Conclusion}
%%%%%%%%%%%%%%%%%%%%%%%%%%%%%%%%%%%%

In conclusion, we demonstrate that 2SIO/1STO has predominantly $c$-axis antiferromagnetic moments, establishing the $n$SIO/1STO SL series as viable artificial analogues to the Ruddlesden-Popper crystals. We furthermore reconcile previous contradictory reports by identifying finite octahedral tilting that generates a net canted moment \cite{Kim2017_STSIOtheory,Matsuno2015_SIOSTO}. Hard x-ray RIXS is shown to be sufficiently sensitive to probe the magnetic interactions that stabilize the observed ground state for the first time. For the bilayer 2SIO/1STO, the magnon gap is significantly smaller than that observed in \seven{} and spin-wave based modelling shows that this material is closer to a phase transition between different ground states. Heterostructuring iridates and probing their magnetic interactions with RIXS thus shows how spin orbit coupling can leverage small structural distortions to alter magnetic interactions with potential to realize quantum critical artificial Ruddlesden-Popper phases.

%%%%%%%%%%%%%%%%%%%%%%%%%%%%%%%%%%%%
\begin{acknowledgements}
%%%%%%%%%%%%%%%%%%%%%%%%%%%%%%%%%%%%

We would like to acknowledge helpful discussions with G. Jackeli, G. Khaliullin, Wei-Guo Yin, and Yilin Wang and experimental assistance from C. Rouleau, Z. Gai, and J. K. Keum. This material is based upon work supported by the U.S. Department of Energy, Office of Basic Energy Sciences, Early Career Award Program under Award No. 1047478. Work at Brookhaven National Laboratory was supported by the U.S. Department of Energy, Office of Science, Office of Basic Energy Sciences, under Contract No.~DE-SC0012704. J.L. acknowledges the support by the Science Alliance Joint Directed Research \& Development Program and the Transdisciplinary Academy Program at the University of Tennessee. J. L. also acknowledges support by the DOD-DARPA under Grant No. HR0011-16-1-0005. A portion of the fabrication and characterization was conducted at the Center for Nanophase Materials Sciences, which is a DOE Office of Science User Facility. Use of the Advanced Photon Source, an Office of Science User Facility operated for the U.S. DOE, OS by Argonne National Laboratory, was supported by the U.S. DOE under Contract No. DE-AC02-06CH11357. X. Liu acknowledges support by MOST (Grant No.2015CB921302) and CAS (Grant No. XDB07020200).

\end{acknowledgements}

\bibliography{refs}

\end{document}